\documentclass [a4paper, 14pt, dvips] {article}
\usepackage{graphicx}
\usepackage{psfrag}

\begin{document}


\title{\bf Exotic Searches in ATLAS}
         \author{Nenad Vranjes, {\it on behalf of the ATLAS Collaboration} \\
              Institute of Physics, Belgrade\footnote{Now at DSM/IRFU, CEA Saclay}}
\date{}
\maketitle

\begin{abstract} {The latest searches for (non-SUSY) Beyond Standard Model phenomena performed with the 
ATLAS detector are reported. The searches have been performed with data from proton-proton  
collisions at a centre-of-mass energy of 7 TeV collected in 2010 and 2011. Various experimental 
signatures have been studied involving the reconstruction and measurement of leptons, photons, jets,  
missing transverse energy, as well as the reconstruction of top quarks. For most of the signatures, 
the experimental reach is signiÞcantly increased with respect to previous results. }
\end{abstract}

\section{Introduction}

The Standard Model of elementary particles (SM) has had enormous phenomenological success during  
the past half century. However, it is generally believed that it is a low energy limit of a more general 
theory, and numerous theoretical scenarios are studied in order to address at least some of the open  
issues such as the uniÞcation of forces, existence of dark matter or hierarchy problem. Many of these 
models indicate that new phenomena are likely to appear at the TeV scale, the direct probe of which is 
possible with the Large Hadron Collider (LHC) at CERN. Supersymmetry (SUSY) is most commonly  
invoked to address the opened questions of the SM, but there are a large number of important and 
well-motivated theoretical models that seek to answer some or all of fundamental questions. These, 
non-SUSY, phenomena are often collectively referred to as ÕExoticÕ physics. The ATLAS experiment \cite{atlas} 
explores TeV-scale territory performing a large number of searches for BSM physics. A comprehensive 
review is beyond the scope of this document, but for the up to date list of published ATLAS results, 
please consult official ATLAS Exotic page \cite{atlaspub}. The emphasis in this conference report is given on the 
recent results. 

This document is organized as follows. After a brief description of data used by ATLAS, searches 
for two-object resonances reconstructed using leptons, jets, photons and missing energy are described. 
This is followed by a review of the ATLAS searches for Extra Dimensions, including searches for TeV-scale 
gravity (a.k.a. black holes). After that, a brief description of some other exotic searches, like contact 
interactions, the search for right-handed W bosons and Majorana neutrinos, leptoquarks, and exotic top 
partners are described. It should be emphasized that searches for new physics are typically signature 
based, although most analyses do include interpretations in one or more models. A full review of all  
possible experimental signatures is also beyond the scope of this report. 

 \section{The LHC, ATLAS Detector and Data Samples} 
 
\noindent
The searches presented in this report have been performed 
with the data from proton-proton collisions at a centre-of-mass energy of 7 TeV collected in 2010 and 2011 with the ATLAS detector at the LHC.  
Excellent performance of the LHC in 2010, leading to 45 pb$^{-1}$ of data recorded by ATLAS, 
continued also in 2011.  The data taking started with luminosities similar to those at the end of 2010, with instantaneous 
luminosity reaching 3.65$\times$10$^{33}$ cm$^{-2}$s$^{-1}$, and on the Òbest dayÓ 122 pb$^{-1}$ was recorded. 
The total integrated luminosity delivered to ATLAS was more than 5 fb$^{-1}$, 
five times larger than previously targeted for 2011. ATLAS data taking efficiency was 93.5\%, and 
 fraction of good quality data 90-100\% depending on the detector subsystem.
The most striking change in beam conditions between 2010 and 2011 data-taking has been
increase in number of interactions per bunch crossing at a given point in an LHC fill.
With up to 34 collisionsÊ per  bunch  crossing, multiple $pp$ interactions (ÒpileupÓ) in both the triggered beam-crossing (Òin-timeÓ) and ones nearby in time (Òout-of-timeÓ)
posed a significant experimental challenge and much effort on understanding its impact on detector performance with data 
and simulation was needed. The performance of the ATLAS detector, in terms of the lepton reconstruction and identification efficiency, 
lepton and jet energy resolutions and scales, as well as the performance of the measurement of the missing transverse energy
($E^{miss}_{T}$) have been established from the data itself. The details of the ATLAS performance are presented at this conference \cite{perf}.
 The searches presented in this report are based on data samples corresponding to integrated luminosities of up to $\sim$2 fb$^{-1}$.
 
 \section{Searches for Heavy Resonances}

The existence of heavy two-object resonances are predicted by a variety of  BSM models.
The analysis basically  consists of search for a bump in the background mass spectra
which is either small or could be parametrized from the data control samples. 
Relatively simple topologies allow for a robust analysis to be done rapidly.
ATLAS Collaboration has searched for dilepton (including lepton+neutrino), diphoton, dijet and jet+photon resonances.
Diphoton searches are described in the section about Extra Dimensions.

\begin{figure}[h!]
\begin{center}
\includegraphics[width=0.325\textwidth]{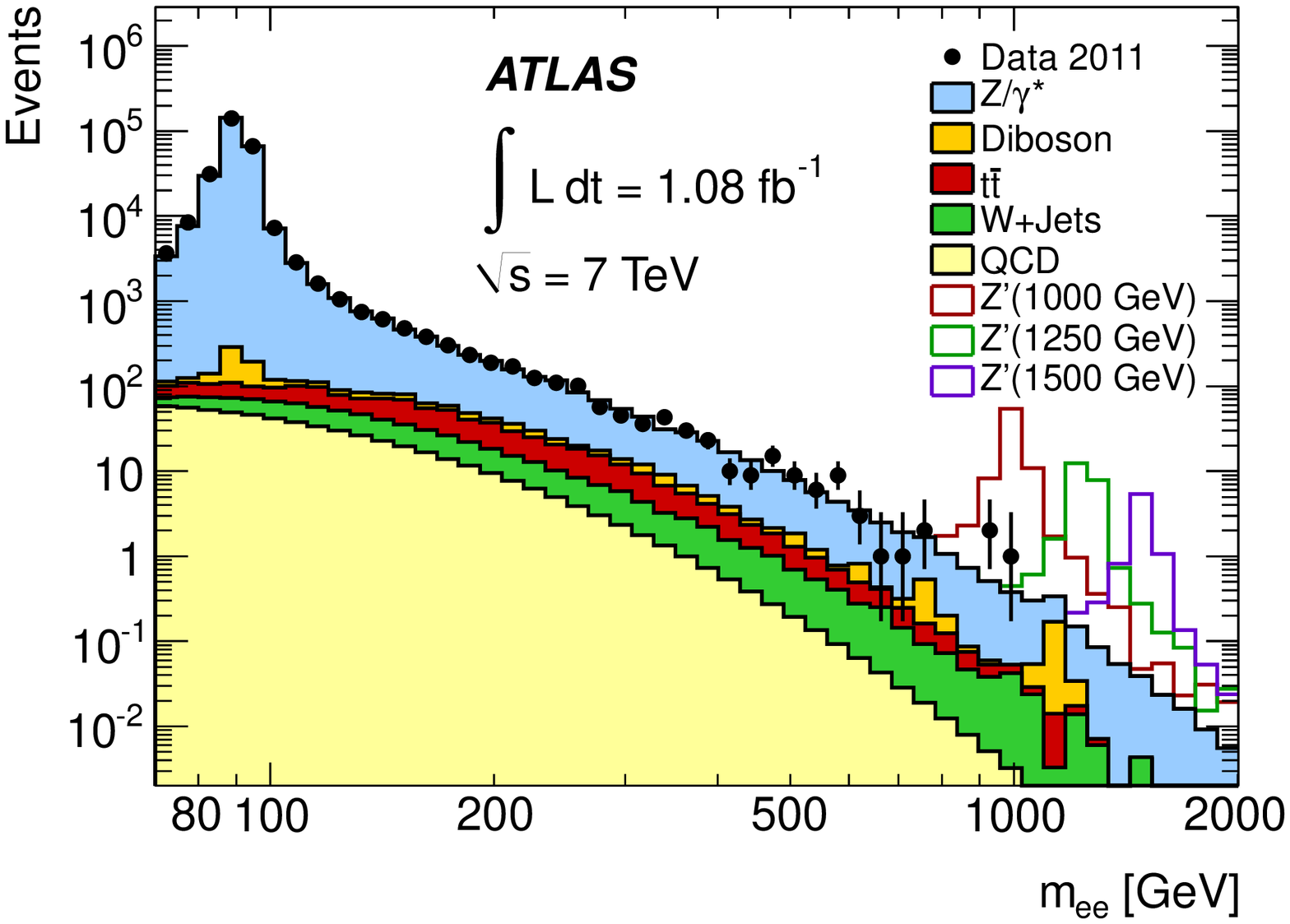}
\includegraphics[width=0.325\textwidth]{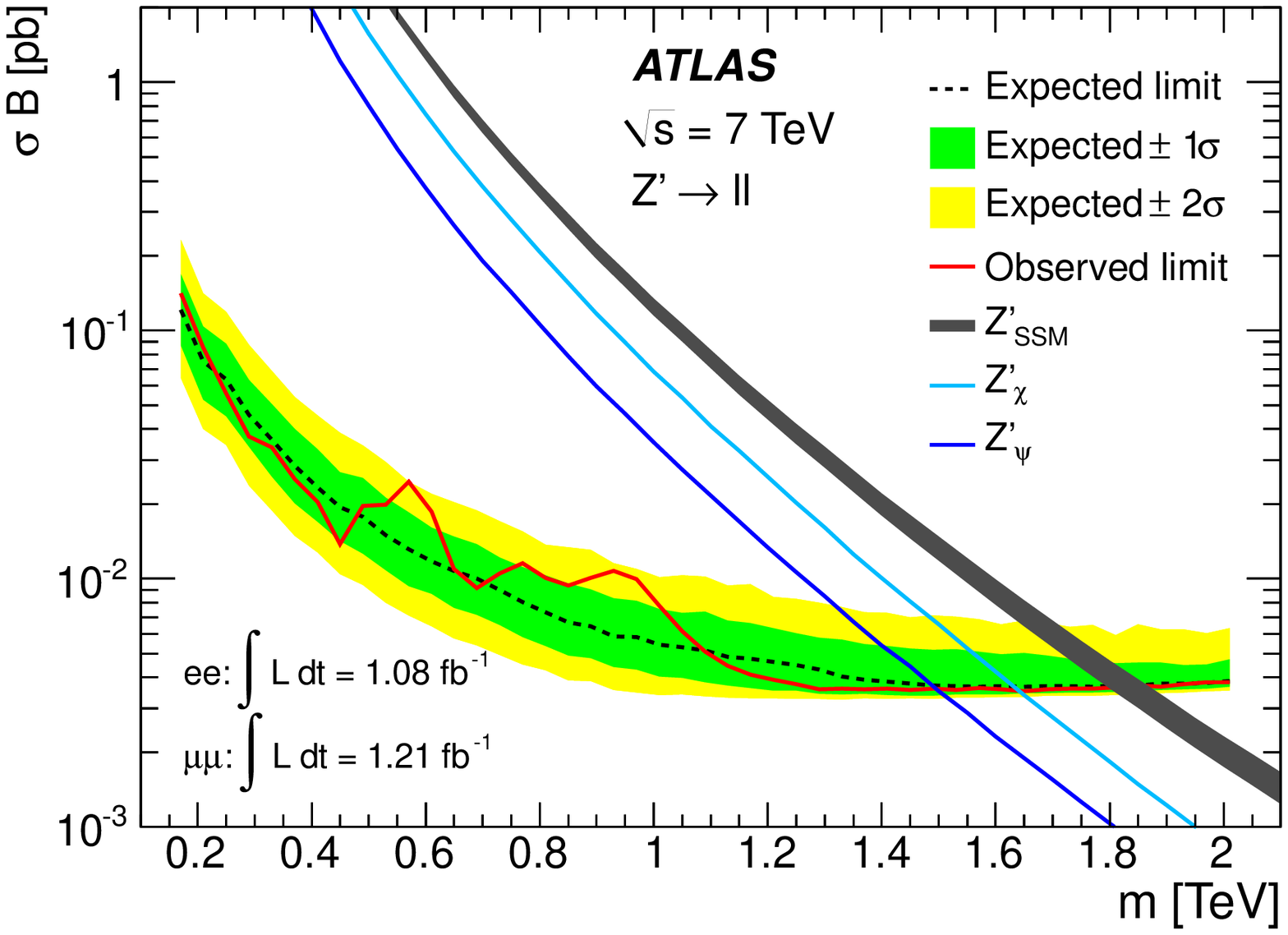}
\includegraphics[width=0.335\textwidth]{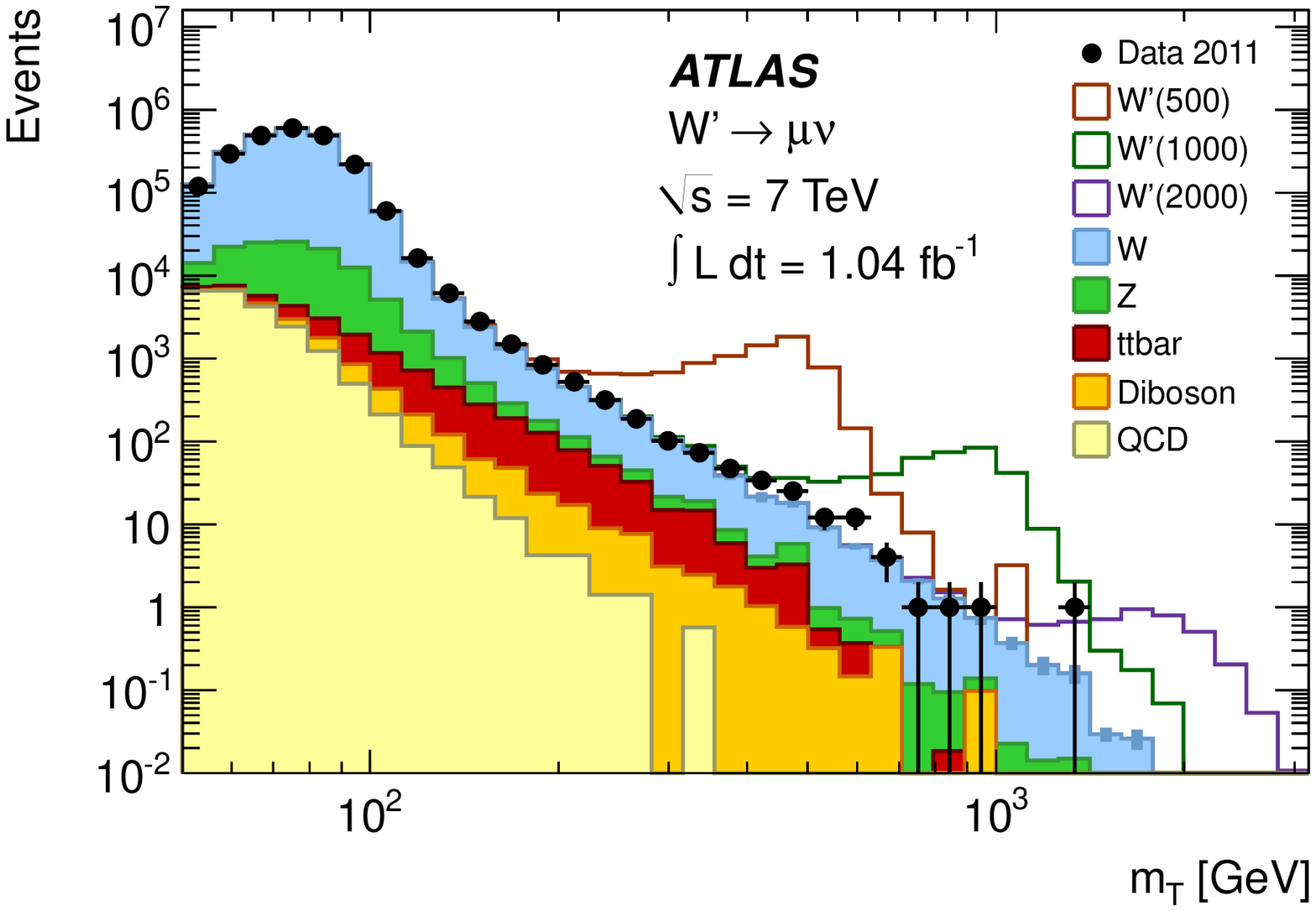}
\end{center}
\caption{Left: Dielectron invariant mass distribution, compared to the expected backgrounds, with three example SSM $Z'$ signals overlaid. Center:
Expected and observed limits on $\sigma B$ and expected $\sigma B$ for SSM $Z'$ production and the two E$_{6}$-motivated $Z'$ models for the combination of $e^{+}e^{-}$ and $\mu^{+}\mu^{-}$ channel. Right: Spectrum of $m_{T}$ for the muon channel in the $W'$ search.}
\label{fig:dileptonfig}
\end{figure}

\noindent
\underline{Dilepton resonances.} Here, a high mass resonance decaying to a pair of isolated leptons $e^{+}e^{-}$, or $\mu^{+}\mu^{-}$ 
is searched for \cite{zprime}. Such resonances include spin-1 heavy neutral gauge bosons such as
sequential  SM-like $Z'$ \cite{zpth}, as well as spin-2 Randall-Sundrum gravitons $G^{*}$ \cite{rsth}. 
The recent ATLAS results is based on the analysis of 1.1-1.2 fb$^{-1}$ of data.
Good lepton reconstruction and identification with high efficiency and resolution under control is crucial for the
searches for high mass dilepton (and especially lepton+neutrino) resonances. For example identification and quality criteria, as well as calorimeter isolation requirements,
are imposed to suppress background from jets faking electrons. To optimize the muon momentum resolution, each muon candidate 
is required to pass quality cuts in the Inner Detector (ID) and to have at least three hits in each of the inner, middle, and outer layers of the Muon Spectrometer (MS). 
The effects of misalignments and intrinsic position resolution are included in the simulation. The muon $p_{T}$ resolution 
at 1 TeV ranges from 15\% (central) to 44\% (for $|\eta|>$ 2). The background is dominated by the irreducible background due to the $Z/\gamma^*$ 
(Drell-Yan) process, characterized by the same final state as the signal. Small contributions come from 
from $t\bar{t}$ and diboson production. QCD background consist of semi-leptonic decays of $b$ and $c$ quarks in the $\mu^{+}\mu^{-}$ channel, and
a mixture of photon conversions, semi-leptonic heavy quark decays, and hadrons faking electrons in the $e^{+}e^{-}$ channel.
Jets accompanying $W$ bosons may similarly produce lepton candidates. Backgrounds from cosmic rays are negligible. 
Level of background, except for QCD background, are initially estimated from Monte Carlo, and after that 
simulated backgrounds are rescaled so that the sum of the background matches the observed number
of events in data around the $Z$ peak. The advantage is that the uncertainty on the luminosity, and any mass independent 
uncertainty on efficiency cancel between $Z'$ and the $Z$ background. The dielectron mass distribution is shown in Fig
\ref{fig:dileptonfig} (left). No significant deviation from SM expectation is observed in either $e^{+}e^{-}$ or $\mu^{+}\mu^{-}$ invariant mass spectra,
and limits\footnote{All limits in this document are at 95\% confidence level (CL).} are set on cross section times branching ratio
($\sigma B$) for a number of $Z'$  models, Fig \ref{fig:dileptonfig} (center). Observed lower mass limit on SM-like $Z'$ is 1.83 TeV
from the combination of both channels and now supersedes Tevatron limit. Observed mass limit on $G^{*}$ is 1.63 TeV, and also suppresses
the current Tevatron result. It is worth noting that the observed upper limits on $\sigma B$ in the wide range of $Z'$ masses between 0.2 TeV to 1.5 TeV
are lower than the Tevatron limits, reflecting increase in new physics sensitivity at the LHC with respect to the Tevatron.
  
Another example of search for dilepton resonances, is the search for technivector mesons $\rho_{T}$ and $\omega_{T}$ \cite{technicolor}.
Technicolor (TC) models predict new technihadron states that could be copiously produced at the LHC. 
The lowest mass states are the scalar technipion $\pi^{\pm,0}$ and the vector technirho $\pi^{\pm,0}$ and techniomega $\omega^0_T$. 
Additional motivation for this search stems from the observation by CDF of an excess in the dijet 
mass spectrum in $Wjj$ events, which has led to speculation that the excess may be due to a 290 GeV $\rho_{T}$  that decays into a $W$ and a 160 GeV $\pi_{T}$.
Since the neutral technimesons are narrow, spin 1 resonances, the search methodology is identical to that 
developed for the $Z'\rightarrow l^{+}l^{-}$ search. TC mesons are excluded for $\rho_{T}$ masses 130 - 480 GeV and for $\pi_{T}$ masses 50-480 GeV.

\noindent
\underline{Lepton+neutrino resonances.} As for heavy neutral $Z'$, many models predict additional heavy charged gauge bosons,  commonly denoted $W'$.
The experimental signature consist of exactly one isolated lepton (electron or muon), and
high $E^{miss}_{T}$ arising from neutrino escaping detection. Since the longitudinal component of the neutrino's 
momentum cannot be unambitiously reconstructed,  transverse mass variable defined as $m_{T} = \sqrt{p_{T}E^{miss}_{T} (1 - cos\phi_{l\nu})}$
is used as a discriminant variable. The most recent ATLAS  results are based on the analysis of $pp$ collisions corresponding to an 
integrated luminosity of 1.04 fb$^{-1}$ \cite{wprime}. The main background to the $W'\rightarrow l\nu$ signal comes from the high-$m_{T}$ tail of SM $W$ boson. 
This, as well as $Z\rightarrow ll$, top and diboson backgrounds are estimated using
Monte Carlo simulation, while QCD and cosmic backgrounds are estimated using data driven techniques.
Transverse mass in the muon channel is shown in Fig \ref{fig:dileptonfig} (left).
No excess  beyond Standard Model expectations is observed. $W'$ with SM-like couplings is excluded at the 95\% CL for masses up to 2.15 TeV
using both $W'\rightarrow e\nu$ and $W'\rightarrow \mu\nu$ channels. The limit on $\sigma B$ suppress the Tevatron limits for $m_{W'}>$ 600 GeV.
The obtained lower mass limit represents an increase of about 50\% (approximately 700 GeV) 
with respect to the mass limit obtained by analyzing 36 pb$^{-1}$ of 2010 data \cite{wprime1}.

\begin{figure}[h!]
\begin{center}
\includegraphics[width=0.35\textwidth]{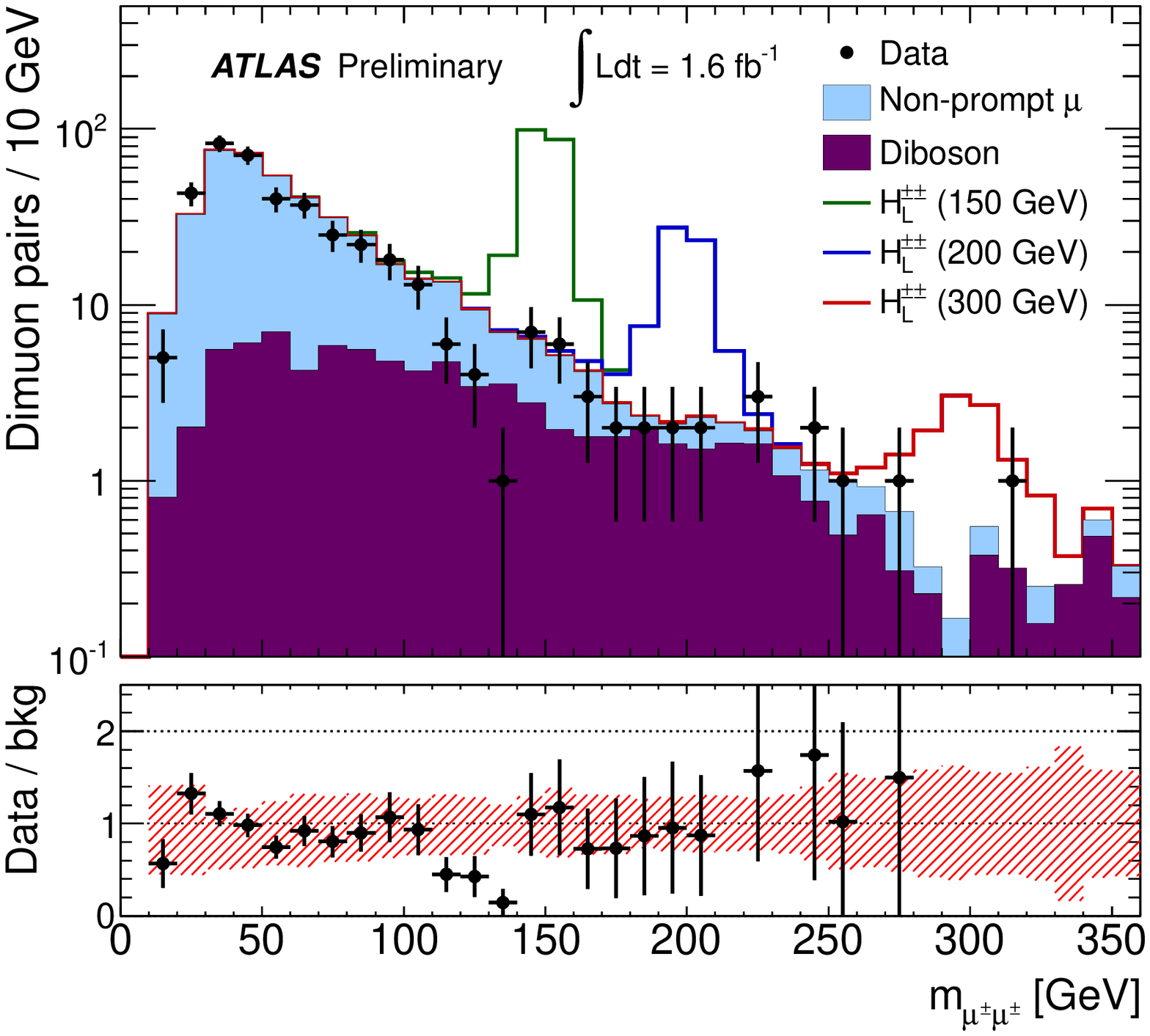}
\includegraphics[width=0.31\textwidth]{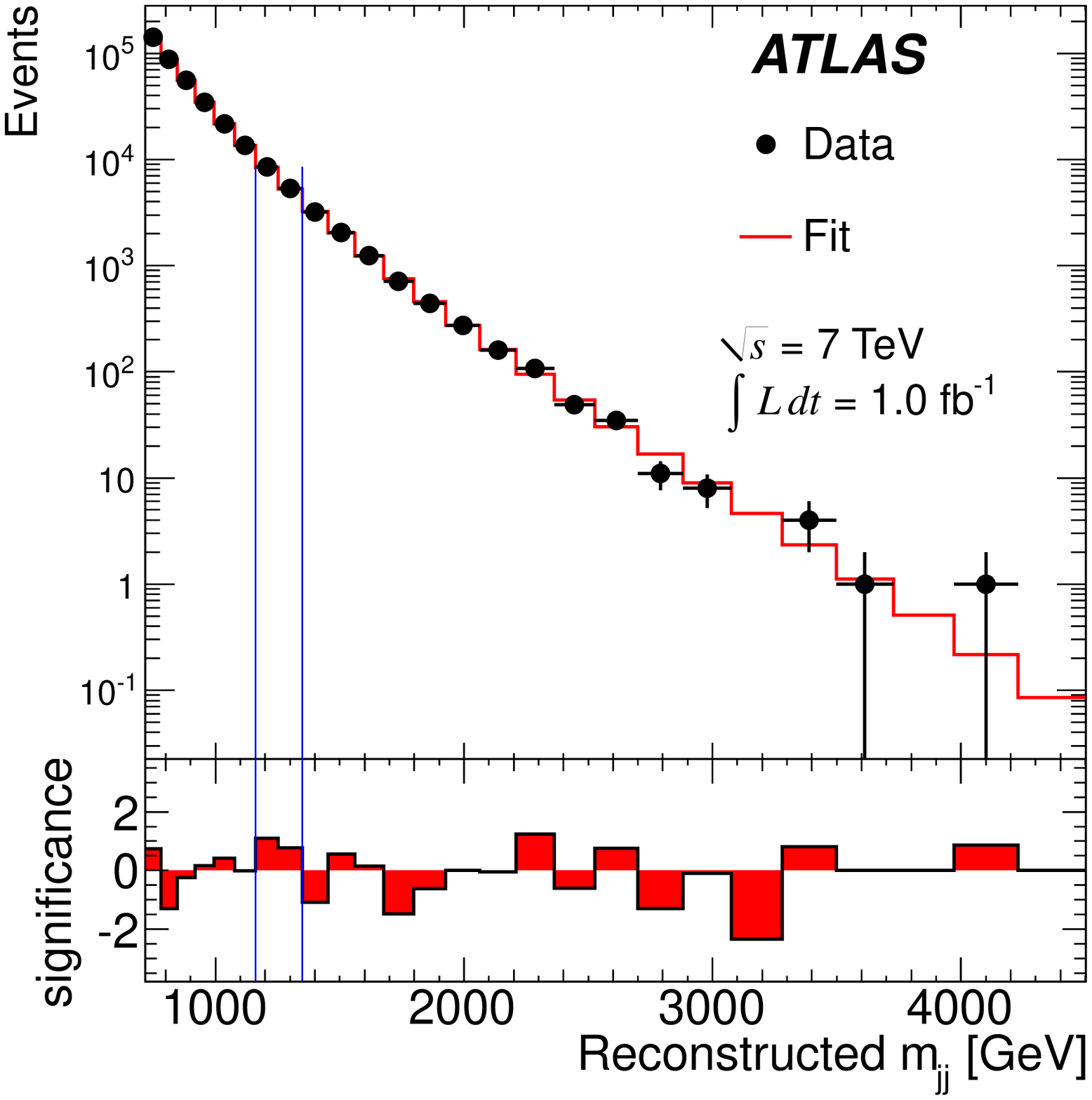}
\includegraphics[width=0.31\textwidth]{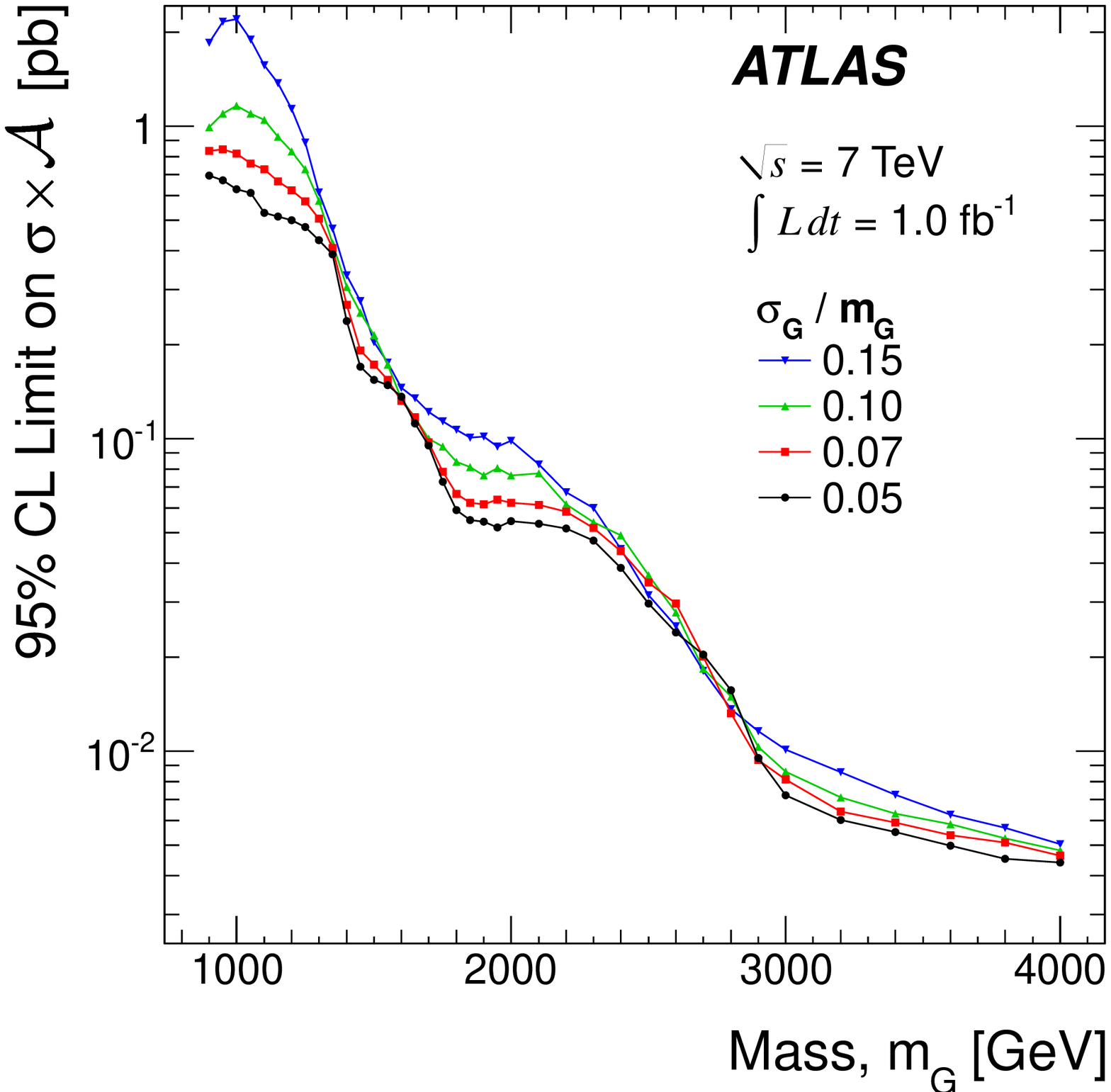}
\end{center}
\caption{Left: Invariant mass distribution for $\mu^{\pm}\mu^{\pm}$  and for hypothetical left-handed $H^{\pm\pm}$. Center: The dijet mass distribution fitted with a smooth functional form describing the QCD background. Right: The upper limits on $\sigma \times A$ for a dijet Gaussian resonance $m_{G}$, with four values of $\sigma_{G}/m_{G}$.}
\label{fig:dimudijet}
\end{figure}

\noindent
\underline{Same sign dimuon resonances.}
Production of prompt same sign muon pairs is predicted by several BSM theories (SUSY, Universal Extra Dimensions, 
heavy Majorana neutrinos, fourth-generation quarks and doubly charged Higgs bosons).
This production, however,  occurs comparatively rarely in the Standard Model. 
Here  the ATLAS search in the context of the doubly charged Higgs boson ($H^{\pm\pm}\rightarrow \mu^{\pm}\mu^{\pm}$) is presented \cite{ssdimuons}.
The fiducial region is defined as two like-sign muons with $|\eta|<$2.5 and $p_{T}>$ 20 (10) GeV
for the leading (subleading) muon that are outside hadronic jets, and have an invariant mass $m_{\mu\mu}>$15 GeV. This fiducial region corresponds well to the experimental 
requirements made in the event selection. In order to suppress background from non-prompt muon production and cosmic
muons, the transverse and longitudinal impact parameters (measured with respect to the event primary vertex) of the muon candidates must 
be small. Muons are required to be isolated from other tracks in ID, and sign of the muon track
is required to be the same in ID and MS. Background is composed of the 
events coming from diboson production ($WZ,ZZ$) and non-prompt 
muon production (semi-leptonic $b$ and $c$ decays, $\pi/K$ decays, cosmic rays).
Non-prompt muon background is estimated using several data-driven techniques, and good agreement between data and
background estimation is obtained, as demonstrated in Fig\ref{fig:dimudijet} (left).
Lower mass limits of  375(295) GeV are set on left-handed (right-handed) doubly charged Higgs production 
assuming the branching ratio to muons is 100\%. These are most stringent limit to date.
Limits are calculated in the mass - branching ratio plane as well. Further, model independent upper limits 
on the cross section of new physics contributing to like-sign dimuon production are placed as function 
of the lower dimuon mass cut assuming the same acceptance: the limit ranges from 304 fb for $m_{\mu\mu}>$ 15 GeV to 5.0 fb for $m_{\mu\mu}>$ 300 GeV. 
The  limits obtained represent a significant improvement compared to those obtained from the analysis based on the 2010. 

\noindent
\underline{Dijet resonances.}
Another search is the hunt for a bump in the dijet mass spectrum. The main observable is $m_{jj}$,  the mass of the system of  the two leading jets in $p_{T}$. 
Studies have shown that the dijet mass distribution are sensitive to the highest mass scales accessible in hadronic final states. Dijet mass distribution is examined in a search for resonances due to new phenomena localized near a given mass, employing a data driven background estimation that does not rely
on detailed QCD calculations. ATLAS Collaboration has receantly updated the search for new physics in the dijet mass 
distribution using 1 fb$^{-1}$ of 2011 data  \cite{dijetresonances}. Jets are reconstructed using the anti-$k_{t}$ algorithm, with the distance parameter $R=$0.6, with momentum calibration based on MC studies including full  detector simulation, and validated with extensive test-beam  and collision data studies.
Measured dijet mass distributions are not corrected for detector resolution, which, in terms of mass smearing, is 
$\frac{\sigma_{m_{jj}}}{m_{jj}}$=5\% at $m_{jj}$=1 TeV, dropping to 4.5\% at 2 TeV, and asymptotically approaching 4\% at $m_{jj}$ of 5 TeV and above.
For final event selection, additional kinematic criteria are applied, requiring that  the two leading jets have $|\eta_{j}|<$2.8, and  the rapidity in the parton CM frame satisfies 
$|y^{*}|<$0.6, where $y$ is the jet rapidity defined as $y\equiv \frac{1}{2}\frac{E+p_{z}}{E-{p_{z}}}$
with $E$ being energy, and $p_{z}$ $z$-component of the jet's momentum, and $y^{*}=\frac{1}{2}(y_{1}-y_{2})$.
These criteria favour central collisions and have been  shown, based on studies of expected signals and QCD 
background, to optimise the analysis sensitivity. The observed dijet mass distribution after all selection 
cuts is shown in Fig.\ref{fig:dimudijet} (center), with highest entry $>$ 4 TeV. $m_{jj}$ spectrum is fitted to the smooth functional form which has been 
empirically shown to fit well data as well as the QCD prediction given by several MC programs.
Limits are derived on cross section times acceptance ($\sigma \times A$) on several new physics models, among which excited quarks are excluded 
bellow masses of 2.99 TeV. Fig. \ref{fig:dimudijet} (right) shows model-independent limits on $\sigma \times A$ for
a collection of hypothetical signals that are assumed to be Gaussian-distributed in $m_{jj}$ with mean ($m_{G}$) ranging 
from 0.9 to 4.0 TeV and standard deviation ($\sigma_{G}$) from  5\% to 15\% of the mean.

\noindent
\underline{$\gamma +$ jet resonances.}  Analogue to the dijet resonance search, a search for a resonance consisting of a
back-to-back jet and a isolated photon is performed with the ATLAS detector \cite{gamajetresonances}.
The photon-jet invariant mass ($m_{\gamma j}$) distribution is used as a main discriminant variable. Background 
resulting from the mixture of direct photon-jet production, from radiation of final-state quarks or from multijet
processes where dijet or higher-order events produce secondary photons during fragmentation of the hard- 
scatter quarks and gluons, is estimated by an empirical fit in the control region. Limits are set on $\sigma B\times A$ for several exotic models, excluding 
excited-quark model for masses up to 2.46 TeV. Limits are also set on a generic signal 
with Gaussian distribution and arbitrary production cross section, assuming the signal lineshape to be a 
Gaussian distribution with one of three  widths, $\frac{\sigma _{G}}{m_{G}}$=5\%, 7\%, and 10\%.

 \section{Searches for Extra Dimensions}

\noindent
In order to  address hierarchy problem  theories with extra spatial dimensions have been developed.
Also, the complete unification of particle forces may require the existence of the additional spacial dimensions, 
three of which are within of our senses, and others being compactified at distance of the order of 10$^{-32}$ m. 
These scenarios provided exciting new ground for experimental probes, and some of the recent searches
performed with ATLAS are reported in this section. 

\noindent
\underline{Graviton dilepton and diphoton searches.} The Randall-Sundrum (RS) model of extra dimensions predicts excited Kaluza-Klein (KK) modes of the graviton, which 
appear as spin-2 resonances. These modes have a narrow intrinsic width when $k/\overline{M}_{Pl}<$0.1, where $k$ is 
the spacetime curvature in the extra dimension, and $\overline{M}_{Pl}=M_{Pl}/\sqrt{8\pi}$ is the reduced Planck scale.
KK gravitons in this model would have a mass splitting of order 1 TeV and would appear as new resonances. The 
phenomenology can be described in terms of the mass of  the lightest KK graviton excitation ($m_{G}$) and the dimensionless coupling to the SM fields, 
$k/\overline{M}_{Pl}$. It is theoretically preferred for $k/\overline{M}_{Pl}$ to have a value in the range from 0.01 to 0.1. 
The recent ATLAS search for dilepton final states \cite{zprime} is described earlier.
The diphoton final state provides a sensitive channel for the search for extra dimensions due to the clean experimental signature, excellent diphoton mass resolution, and modest backgrounds,  as well as a branching ratio for graviton decay to diphotons that is twice the value of that for graviton decay to 
any individual charged-lepton pair. Another model that could be probed with the diphoton final states, is the Arkani-Hamed,    
Dimopoulos and Dvali (ADD) model \cite{addth}. The ADD model postulates the existence of flat additional spatial dimensions compactified with radius $R$, in which only
gravity propagates. The mass splitting between subsequent KK states is  of order $1/R$. In this model, resolving the hierarchy 
problem requires typically small values of $1/R$, giving rise  to an almost continuous spectrum of KK graviton states.
Effects due to ADD graviton exchange would be evidenced by a non-resonant deviation from the SM background expectation.
\begin{figure}[h!]
\begin{center}
\includegraphics[width=0.4\textwidth]{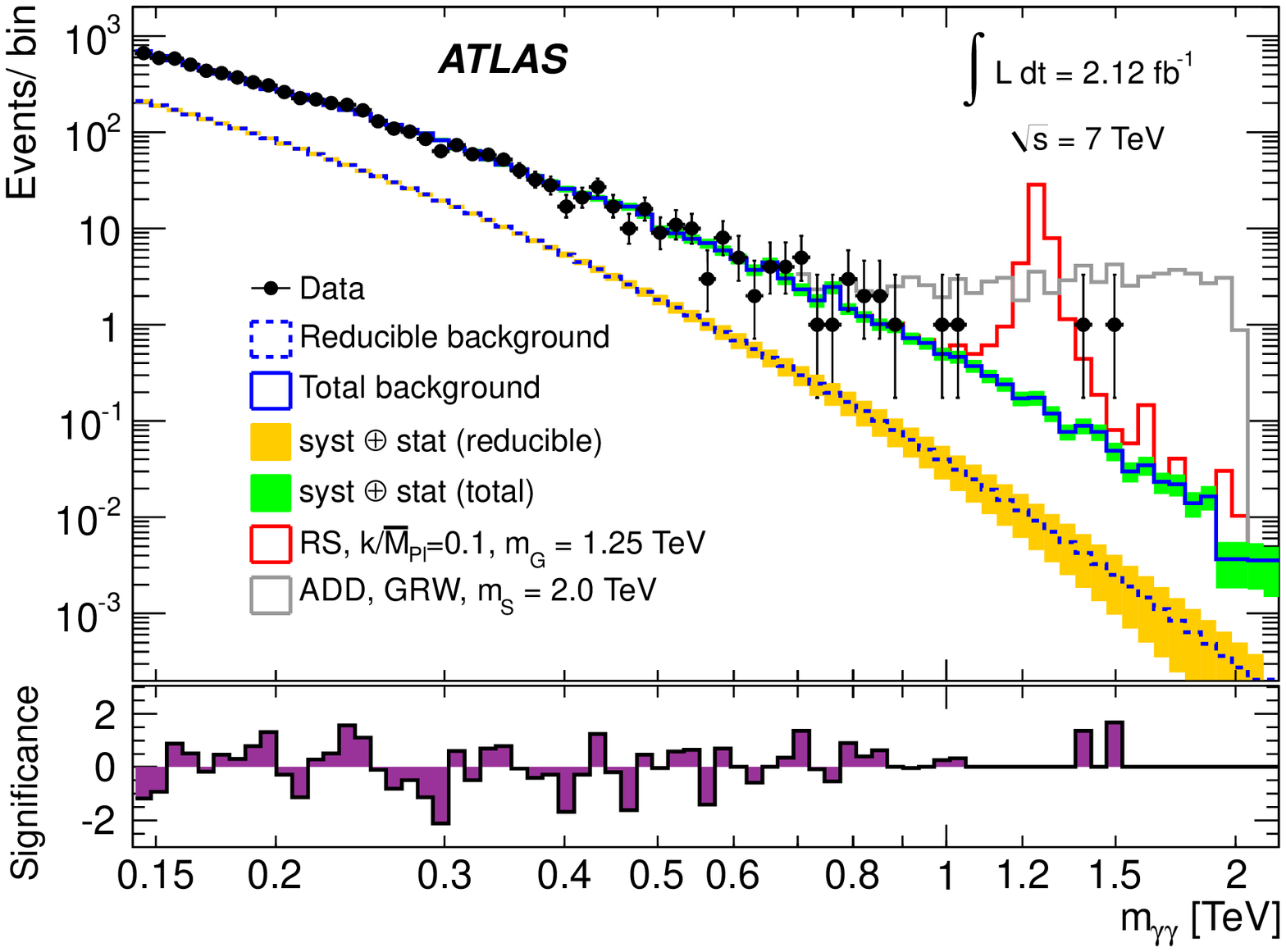}
\includegraphics[width=0.4\textwidth]{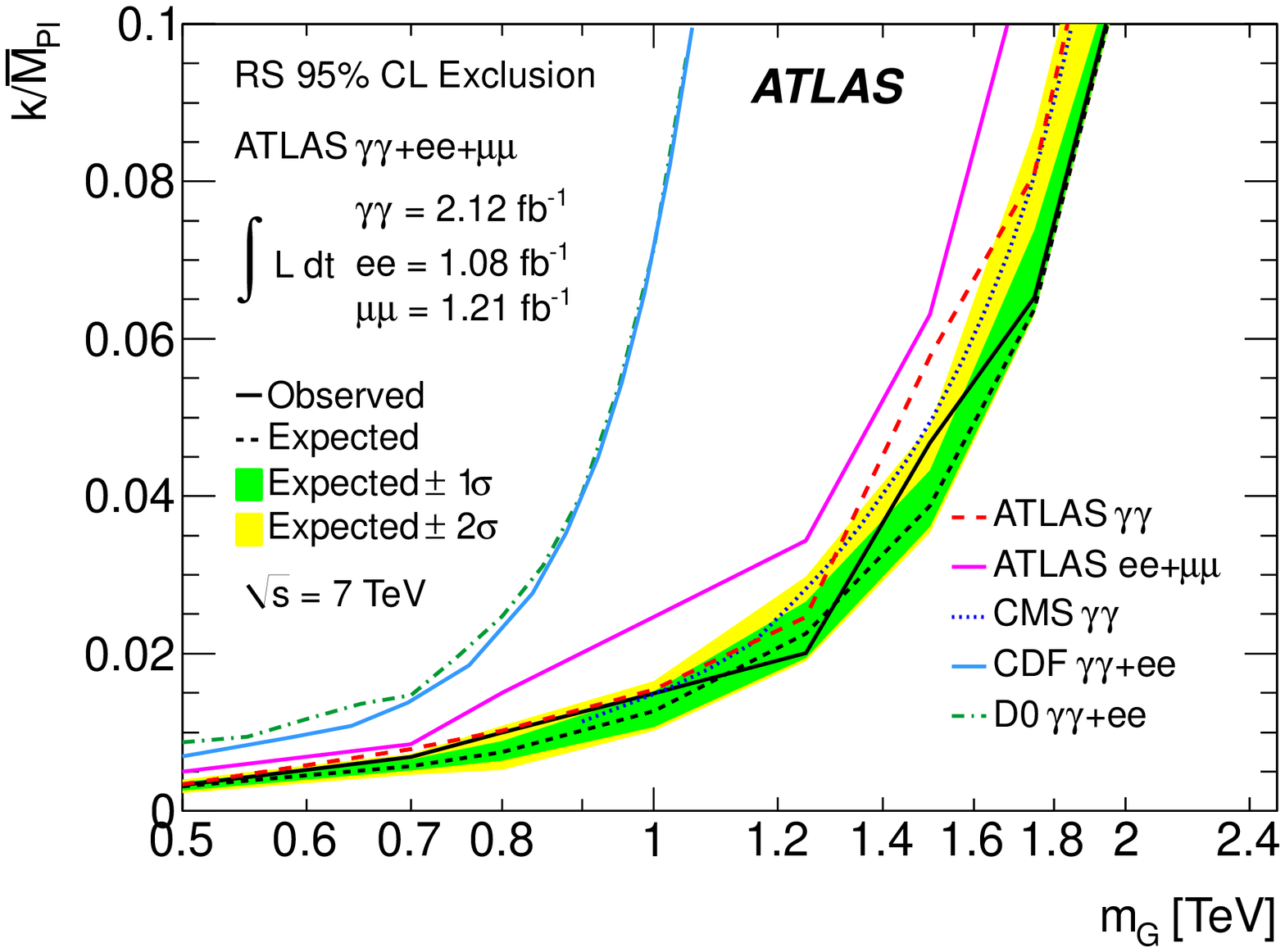}
\end{center}
\caption{Left: The observed $m_{\gamma\gamma}$ distribution of diphoton events, superimposed with the predicted SM background and expected signals for ADD and RS models with indicated choices of parameters. Right: The RS limit interpreted in the plane of coupling versus graviton mass. }
\label{fig:gamgam}
\end{figure}

ATLAS searches for extra dimensions with $\gamma\gamma$ final states performed with 2.12 fb$^{-1}$ of data are described in \cite{EDgamagama}.
The largest background for this analysis is the irreducible background due to SM $\gamma\gamma$ production. 
The shape of the diphoton invariant mass spectrum from this background was estimated using MC, reweighting  
samples to the dfferential cross section predictions of state-of-the-art higher order generator. 
Another significant background component is the reducible background that includes events in which one or 
both of the reconstructed photon candidates result from a different physics object being misidentiÞed as a 
photon. This background is dominated by $\gamma j$ and $jj$ events, with one or two jets faking photons, 
respectively. This background is estimated using data-driven techniques. Backgrounds with electrons faking photons, such 
as the Drell-Yan production of electron-positron pairs as  well as $W/Z + \gamma$ and top processes, were verified to be 
small.

The resulting $\gamma\gamma$ spectrum is presented in Fig \ref{fig:gamgam} (left). No signal is observed, and
upper limits are determined on the RS and ADD signal cross sections. The RS model results are combined with the limits
obtained from the search with dilepton states, Fig \ref{fig:gamgam} (center). The mass of the lightest RS 
graviton is excluded bellow the value of 1.95 TeV for $k/\overline{M}_{Pl}$=0.1. 
The observed  upper limit on ADD model is 2.49 fb for the product of the cross section due to new physics multiplied by the acceptance and efficiency.

\noindent
\underline{Universal Extra Dimensions: $\gamma\gamma + E^{miss}_{T}$ signature.} 
This analysis considers the case of a single Universal Extra Dimension (UED) \cite{uedth}, with compactification radius
$R \approx$ 1 TeV$^{-1}$.  At the LHC, the main UED process would be the production via the strong interaction of a pair of first-level KK 
quarks and/or gluons. These would decay via cascades  involving other KK particles until reaching the lightest KK 
particle (LKP), i.e. the first level KK photon $\gamma^{*}$ decaying  via $\gamma^{*}\rightarrow \gamma + G$,
where is $G$ a tower of  eV-spaced graviton states, leading to a graviton mass between 0 and $1/R$. With two decay chains per event, the 
final state would contain $\gamma\gamma + E^{miss}_{T}$, where $E^{miss}_{T}$ results from the escaping gravitons.
A search for UED with this event topology has been performed using 1.07 fb$^{-1}$  \cite{Aad:2011zj}.
No excess of events is observed above the Standard Model prediction and upper limits are set on the production cross section for new physics, leading to the 
lower limit of 1.23 TeV  on $1/R$. The results are also interpreted in the context of a generalised model of
gauge-mediated supersymmetry breaking with a bino-like lightest neutralino \cite{ggmet1}, 
and in the context of a minimal model of gauge-mediated supersymmetry breaking \cite{ggmet2}.

\noindent
\underline{Search for monojets.} 
In the ADD Large Extra Dimensions (LED) model, the four-dimensional  Planck scale, $M_{Pl}$, is related to the fundamental 
$4+n$-dimensional Planck scale, $M_{D}$, by $M_{Pl}^{2} \sim M_{D}^{2+n}R^{n}$,  where $n$ and $R$ are the number 
and compactification radius of the extra dimensions, respectively. An appropriate choice of $R$ for a given $n$ allows for a value of 
$M_{D}$ close to the electroweak scale. The compactification of the extra spatial dimensions results in a KK
tower of massive graviton modes. These graviton modes are produced in association with a jet and do not interact with the detector, which 
results in a monojet signature in the final state. ATLAS has published an update of the previous analysis using a much larger 2011 data sample 
corresponding to 1.0 fb$^{-1}$  \cite{monojets}.  Event topology consist of a high energy jet, and large $E^{miss}_{T}$.
Three signal regions with increasing cuts on the jet $p_{T,j}$ and $E^{miss}_{T}$  are defined  {\it i)}'low $p_{T}$' $p_{T,j}>120$ GeV and $E^{miss}_{T}>220$ GeV,  
 {\it ii)} 'high $p_{T}$' $p_{T,j}>250$ GeV and $E^{miss}_{T}>250$ GeV,  {\it iii)}'very high $p_{T}$' $p_{T,j}>350$ GeV and $E^{miss}_{T}>300$ GeV.
 The aim is to maintain the sensitivity to a varity of models for new phenomena.
Jets are required to be central $|\eta_{j}|<$2, and events are vetoed if there is another jet or lepton present in the event.
The expected background to the monojet signature is dominated by $Z \rightarrow \nu\bar{\nu}$+jets and $W$+jets production, 
and is estimated from data itself. The agreement between the data and the SM predictions for the total number of events in the different 
analyses is translated into model-independent  upper limits on the $\sigma \times A$.
The limits are set also in the context of the ADD model: for number of extra dimensions $n$=2-6, $M_{D}>$3.2 TeV - 2.0 TeV.
\begin{figure}[h!]
\begin{center}
\includegraphics[width=0.37\textwidth]{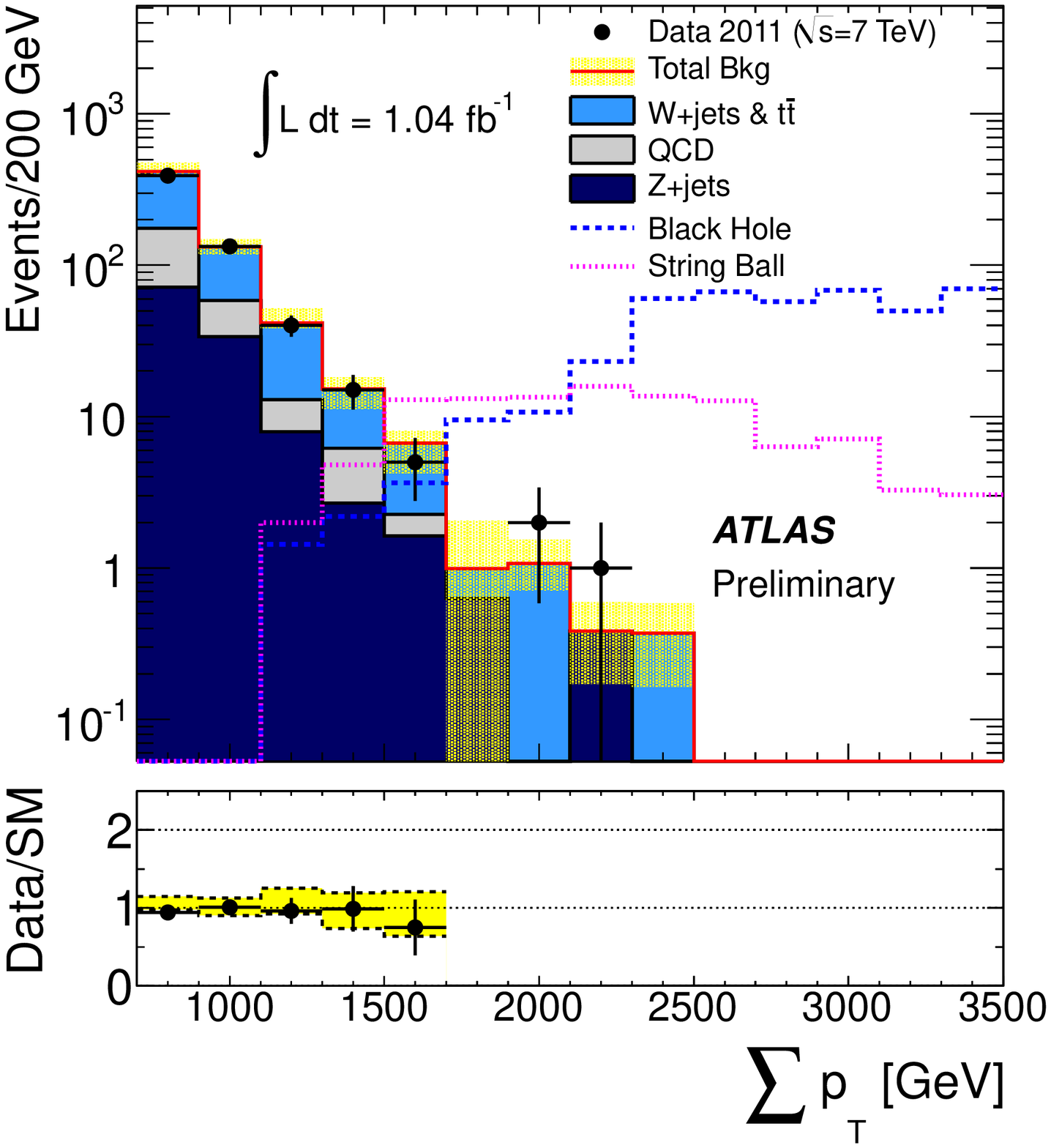}
\includegraphics[width=0.38\textwidth]{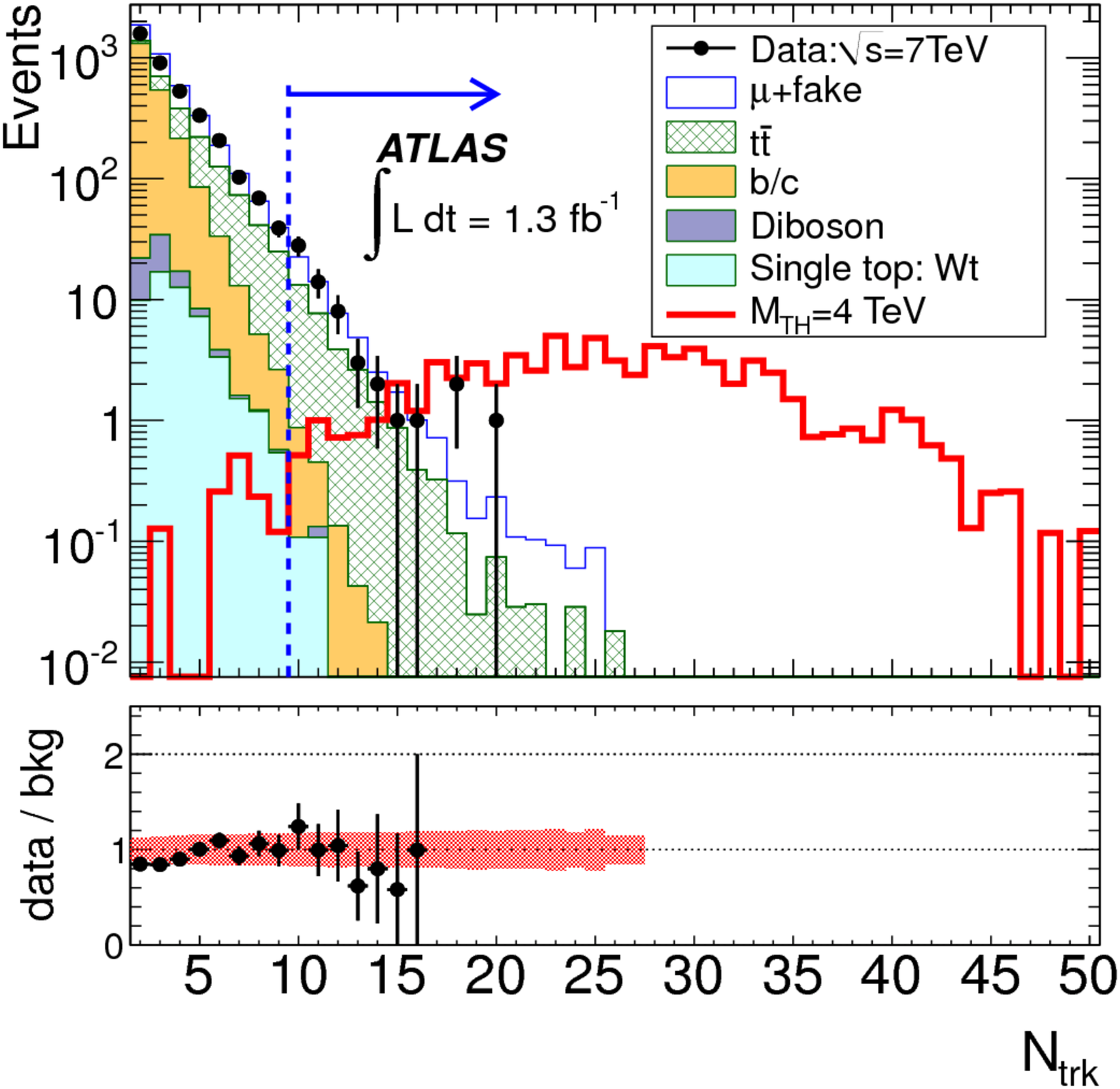}
\end{center}
\caption{Left: $\Sigma p_{T}$ distributions for the signal region in the electron channel of the lepton+jets black hole search. Right: The track multiplicity distribution for same-sign dimuon events. }
\label{fig:bholes}
\end{figure}

\noindent
\underline{Searches for TeV-gravity signature in final states with leptons and jets.}  Low-scale gravity models allow the existence of non-perturbative gravitational states such as black holes. In ADD model black holes can form ÊifÊ $\sqrt{s}>M_{D}$,  andÊ collisionÊ impactÊ parameterÊ isÊ smallerÊ thanÊ higherÊ  dimensional ÊSchwarzschildÊ radius. 
Microscopic Êblack  holesÊ decaying ÊviaÊ HawkingÊ radiationÊ withÊ isotropicÊ and ÊÔdemocraticÕÊ  decayÊ leading ÊtoÊ largeÊ
multiplicityÊ ofÊ particles. Searches for these states have previously been performed by investigating Þnal states with multiple 
high-$p_{T}$ objects with high-$p_{T}$ jets only. Depending on the model 15 - 50\%Ê ofÊ eventsÊ withÊ Êblack  holes  producedÊÊÊÊ 
should  haveÊ Êat Êleast Ê1Ê chargedÊ lepton.  Requiring  the presence of the lepton dramatically ÊreducesÊÊ backgroundsÊ andÊ 
systematic Êuncertainties.  ATLAS  search  for  black  holes in jets + lepton final states is presented here \cite{bhljet}.
Scalar sum of $p_{T}$ of high-$p_{T}$ leptons and jets in the event is used as discriminant variable as shown in Fig \ref{fig:gamgam}.
Model Êindependent ÊlimitsÊ (usingÊ trueÊ fiducialÊ regionÊ separatelyÊ definedÊ forÊ theÊ electronÊ andÊ 
muon Êchannel)  are  calculated  estimatingÊ totalÊ acceptanceÊ of 60 - 90\%Ê 
(40 - 60\%)Ê forÊ electronÊ (muon)Ê channel,  excluding new physics  cross  section 169 fb - 8.7 fbÊÊ(77 fb - 4.8 fb)Ê  
dependingÊ onÊ theÊ cut ÊonÊ $\Sigma p_{T}$.  As an illustration, mass limits within the ADD model are quoted: 
 $M_{D}>$1.0Ê TeV  for ÊÊ$M_{Th} >$4.6Ê TeV, or $M_{D}>$ 2.5Ê TeV,ÊÊ$M_{Th}>$4.0Ê TeV, where $M_{Th}$ is the threshold for the creation of the black hole. Ê 

\noindent
\underline{Search for Strong Gravity Signatures in Same-sign Dimuon Final States.} In this analysis \cite{bhssmumu}, events are selected containing two muons of the 
same charge. This channel is expected to have low Standard Model backgrounds 
while retaining good signal acceptance. In order to maintain optimal acceptance for a possible signal, only one of the 
muons is required to be isolated in this analysis, thereby typically increasing 
the acceptance in the signal region by 50\%.  Black hole 
events typically have a high number of tracks per event ($N_{trk}$), while Standard 
Model processes have sharply falling track multiplicity distributions. So $N_{trk}$ is used as a discriminat variable, and the 
background is estimated from the control region.  No excess of events over 
the SM prediction is observed and exclusion contours are obtained in the plane of the reduced Planck scale $M_{D}$ and $M_{Th}$.
A model independent limit on  $\sigma B \times A$, where $B$ is the branching ratio to dimuons, and $A$ the acceptance of non Standard Model 
contributions,  is derived to be  0.018 pb.

 \section{Searches for Contact Interactions, $W_{R}$ and Majorana Neutrinos, and Leptoquarks}

\begin{figure}[h!]
\begin{center}
\includegraphics[width=0.4\textwidth]{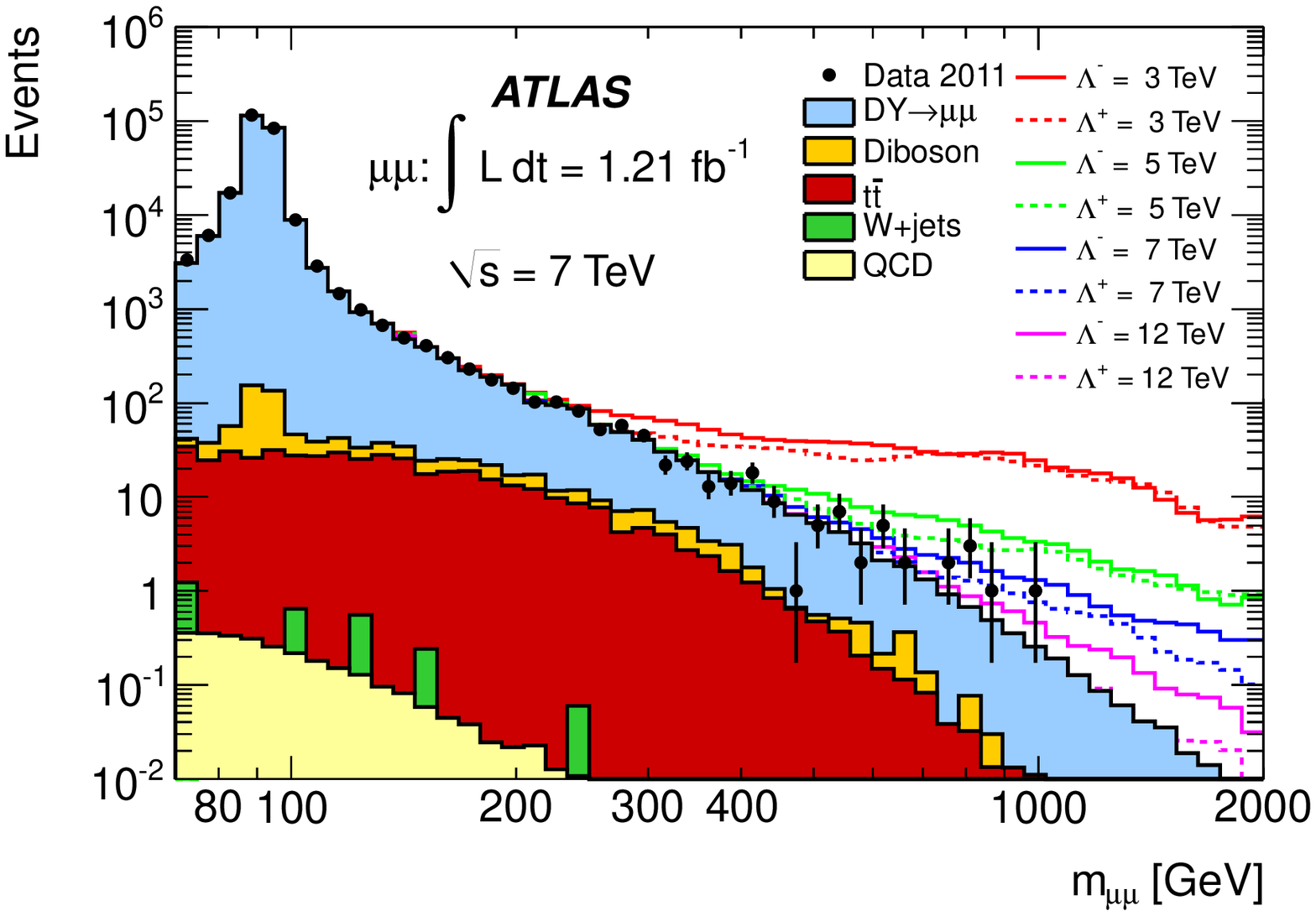}
\includegraphics[width=0.3\textwidth]{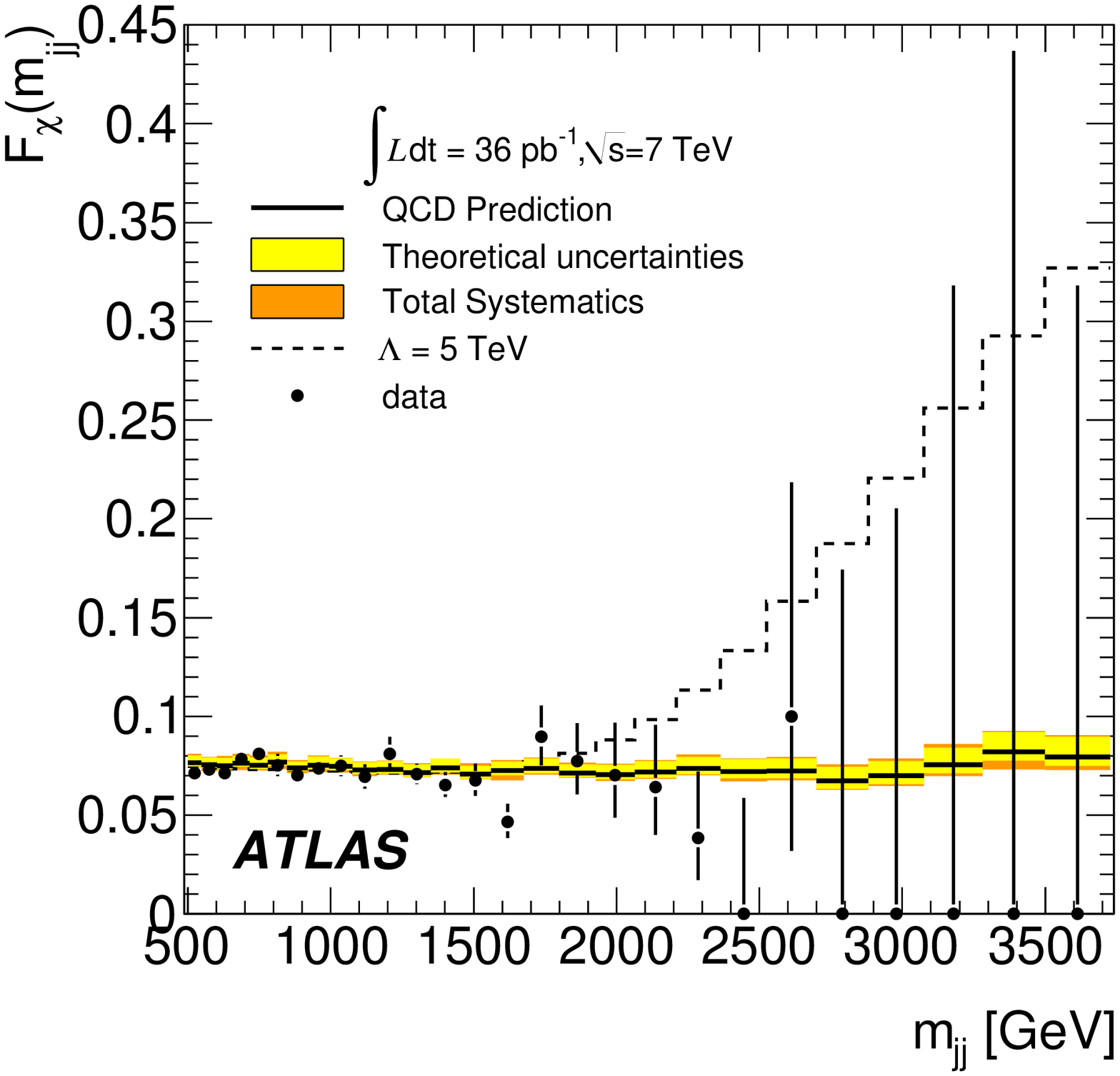}
\end{center}
\caption{Left: $m_{\mu\mu}$ distribution in the contact interaction search. The open histograms correspond to the distributions expected in the presence of contact interactions with different values of $\Lambda$ for constructive or destructive interference. Right: The $F_{\chi}(m_{jj})$ versus $m_{jj}$ and with the expected signal from QCD plus a 
quark contact interaction as described in the text.}
\label{fig:ciqqmmqqqq}
\end{figure}

\underline{Search for Contact interaction.} If quarks and leptons are composite, with at least one common  constituent, the interaction of these constituents would likely be manifested through an effective four-fermion contact interaction (CI) at energies well below the compositeness scale. Such a contact interaction could also describe a new interaction with a messenger too heavy for direct observation at the LHC, in analogy with FermiÕs nuclear $\beta$ decay theory.
A new interaction at high energy scale could be described by the effective Lagrangian, with parameter $\Lambda$ interpreted as a scale bellow 
quarks and leptons constituents are bound. One investigated process is $q\bar{q}l^{+}l^{-}$, $l=e,\mu$ contact interaction \cite{dilepci}. 
The addition of the contact interaction term to the SM Lagrangian describing Drell-Yan production alters the DY production cross  section. The largest  deviations, 
either constructive or destructive, are expected at high dilepton invariant mass and are determined by the scale $\Lambda$.
This analysis interprets the data in the context of the left-left isoscalar model (LLIM), which is commonly used as a benchmark 
for contact interaction searches \cite{cith}. Fig  \ref{fig:ciqqmmqqqq} (left) represents DY mass spectra in the $\mu^{+}\mu^{-}$, with deviation from the spectra if
as a function of $\Lambda$ if constructive/destructive interference had be present. To test the consistency between the data and the 
standard model, a likelihood ratio test was performed. The derived $p$-value  corresponding to the probability of observing a fluctuation corresponds to the likelihood after 
in the pseudoexperiments that is at least as signal-like as  that seen in the data (i.e. with a maximum likelihood ratio greater or equal to that obtained in the data), 
is estimated to be 39\% (79\%) in the electron channel and 21\% (5\%) in the muon channel for constructive (destructive) interference.
These values indicate that there is no significant evidence for contact interactions in the analyzed data and thus limits are set on the contact interaction scale $\Lambda$.  The observed limits are $\Lambda^{-} >$ 10.1 TeV ($\Lambda^{+} >$ 9.4 TeV) in the electron channel and $\Lambda^{-} >$ 8.0 TeV  ($\Lambda^{+} >$ 7.0 TeV) in the muon channel for constructive (destructive) interference.

Another type of contact interaction that has been investigated, albeit with smaller datasample, is $q\bar{q}q\bar{q}$ CI \cite{qqqqci}. Basically, this is dijet signature, 
with event selection similar as in dijet resonant search described previously. The basic observable is $\chi\equiv e^{|\Delta y|}$
where $\Delta y$ is the rapidity difference between the leading and the subleading jet.  A derived 
observable is $F_{\chi}$, which is the fraction of events at $\chi<e^{1.2}$. The choice of $e^{1.2}$ is based on optimization of 
the sensitivity to quark CI, and new physics would appear as an increase in $F_{\chi}$. When $F_{\chi}$ is 
computed in bins of $m_{jj}$, the result is an  spectrum which can indicate new physics by an increase of $F_{\chi}(m_{jj})$ 
in $m_{jj}$ bins that contain significant amounts of signal. Fig. \ref{fig:ciqqmmqqqq} (right) shows the observed and expected $F_{\chi}(m_{jj})$ 
spectra. No such signal is observed, and limit on $\Lambda$ is set to be $\Lambda>$9.5 TeV.

\begin{figure}[h!]
\begin{center}
\includegraphics[width=0.4\textwidth]{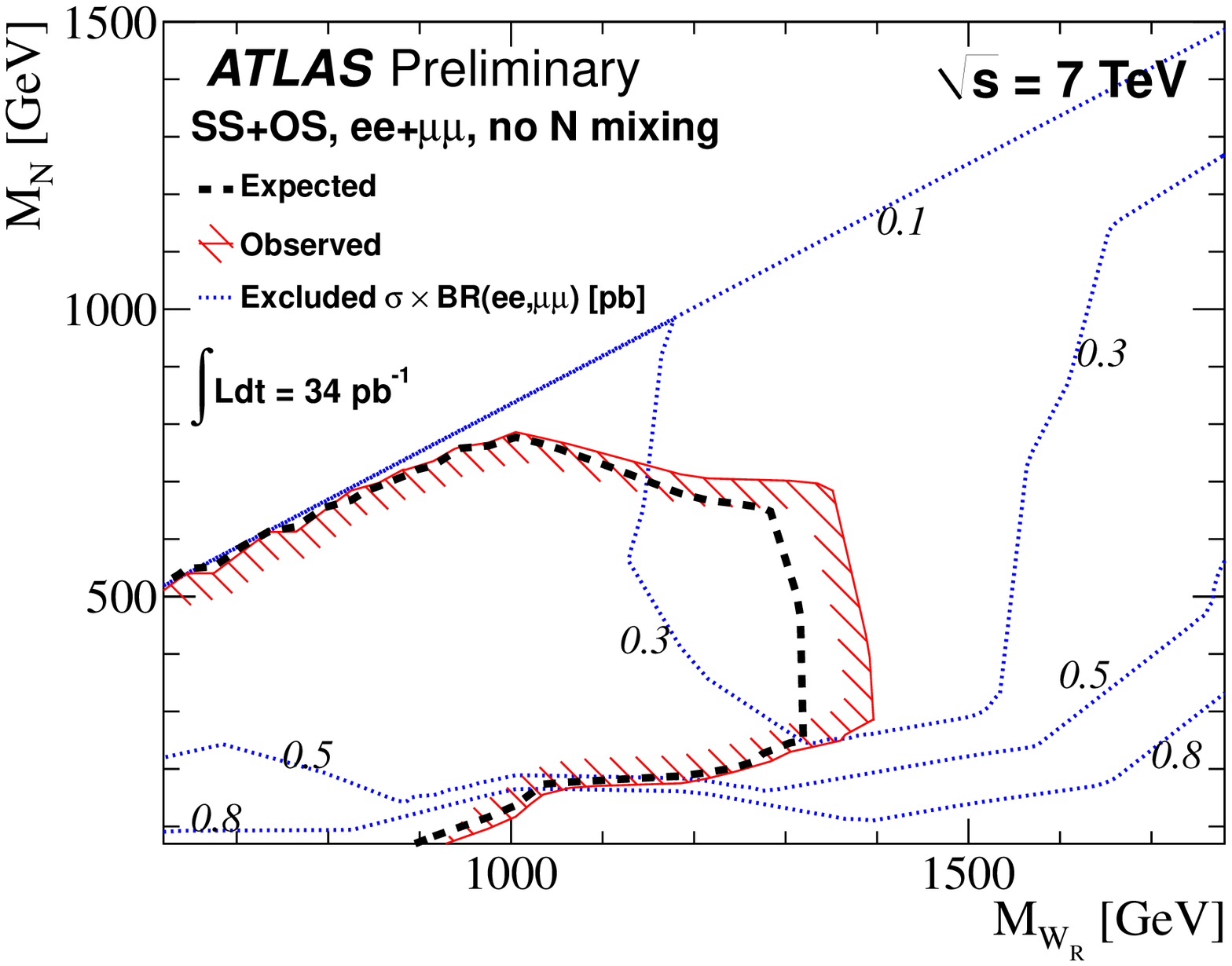}
\includegraphics[width=0.4\textwidth]{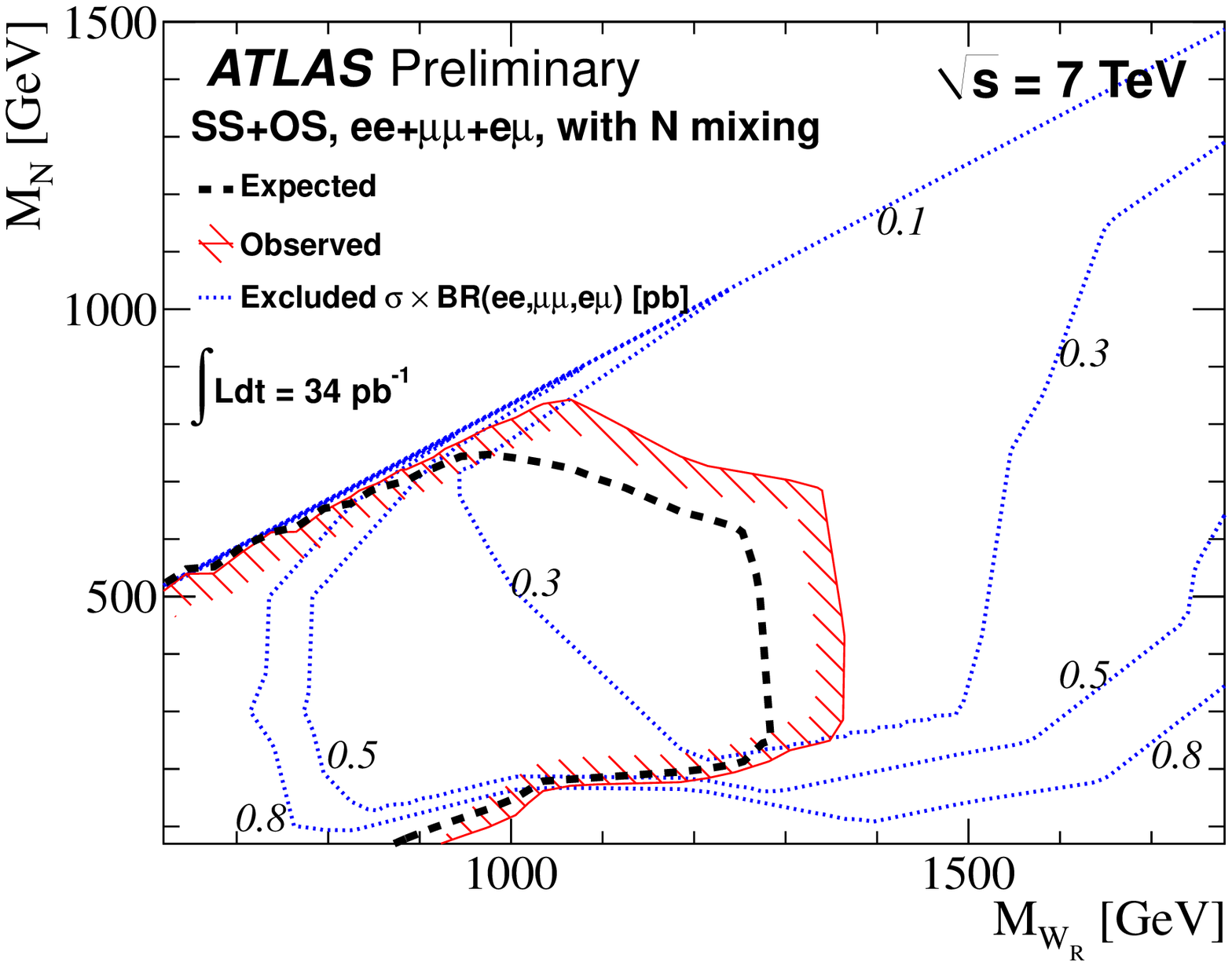}
\end{center}
\caption{Observed and expected limits on the heavy Majorana neutrino and $W_R$ masses for the combined SS+OS channels. The no-mixing scenario is shown (left),
                as well as 100\% mixing scenario (right).}
\label{fig:wrmajo}
\end{figure}

\noindent
\underline{Search for $W_{R}$ and Majorana Neutrinos.} ATLAS has recently released  a search for hypothetical heavy Majorana neutrinos and $W_{R}$ gauge 
bosons  \cite{WRMN}. The results are based on 34 pb$^{-1}$ of data collected in 2010.  The approach is guided by the 
Left-Right Symmetric model (LRSM) \cite{lrsmth} in which a new gauge group with its force particles ($W_{R}$) manifesting themselves directly at LHC energies 
is introduced \cite{lrsmth2}. In this model, the heavy neutrinos  $N$ are produced in the decays of a $W_{R}$ via $q\bar{q} \rightarrow W_{R} \rightarrow lN$, 
with $N$ decaying subsequently to $N \rightarrow lW^{*}_{R} \rightarrow ljj$. 
The final state signature is two high-$p_{T}$ leptons (electron or muon) and two high-$p_{T}$ jets. 
The $N$ and $W_{R}$ invariant masses can be fully reconstructed from the decay products. Similarly 
to SM neutrinos, heavy Majorana neutrinos can mix if their masses are different. Both the scenarios of 
no-mixing and 100\% mixing between two generations of lepton flavour ($e$ or $\mu$) are investigated. 
Both the same-sign (SS) and opposite-sign (OS) dilepton final states are considered. If the heavy neutrinos are of Majorana type, 
they would contribute to both the SS and OS channels, while non-Majorana heavy neutrinos would contribute solely to the OS channel. 
The OS final state can be exploited for the LRSM search since the expected signal events, for the considered $W_{R}$ masses, populate a kinematic 
region where the SM backgrounds are small. The final selection requires $m_{ll}>$110 GeV and $m_{lljj}>$400 GeV in the SS search, while in  OS search
additionally $S_{T}>$400 GeV is required, where $S_{T}$ is the scalar sum of the transverse energies  of the leptons and up to the two leading jets (in $p_{T}$).
The numbers of observed data events and the expected SM background events in the SS and OS final 
states signal regions for the preselection and final selection are in good agreement. The upper limits on the 
$\sigma B$ for new physics based on the LRSM models are obtained. The limits on the heavy Majorana neutrino and $W_{R}$ masses,  for each mixing scenario, are determined for SS events and SS+OS events. The SS+OS limit is obtained by combining the two individual limits, and results are shown in Fig \ref{fig:wrmajo}.

\noindent
\underline{Search for the first generation of leptoquarks.} A search for the first generation of scalar leptoquarks using 1.03 fb$^{-1}$ of $pp$ data has been receantly relased by the ATLAS \cite{lq}. Leptoquarks are sought via their decay into an electron or neutrino and a quark, 
producing events with two oppositely charged electrons and at least two jets, or events with an electron, large $E^{miss}_{T}$
and at least two jets. Control data samples are used to validate background predictions from MC. Likelihood ratio is
used to separate signal from background  in  order  to  allow  mass independent optimization of the search. The  following  discriminating  variables,  selected  to  give  the 
best  separation  between  signal  and  background,  are used  for  the construction  of  the likelihood: {\it i)} for the $eejj$  channel scalar sum of electron and jet $p_{T}$ ($S_{T}$), 
and the average invariant LQ mass;  {\it ii)} for the $e\nu jj$ channel $m_{T}$, $S_{T}$, the transverse LQ mass, and the invariant LQ mass $mLQ (e, jet)$ are used.  No
signal is observed and  limits are set on $M_{LQ}>$ 662 GeV  assuming 100\% branching  ratio to lepton+quark and $M_{LQÊ}>$ 606 GeV
assuming 50\% branching ratio to lepton+quark.

\section{Searches for New Phenomena in $t\bar{t}$ Events with Large Missing Energy}

\begin{figure}[h!]
\begin{center}
\includegraphics[width=0.4\textwidth]{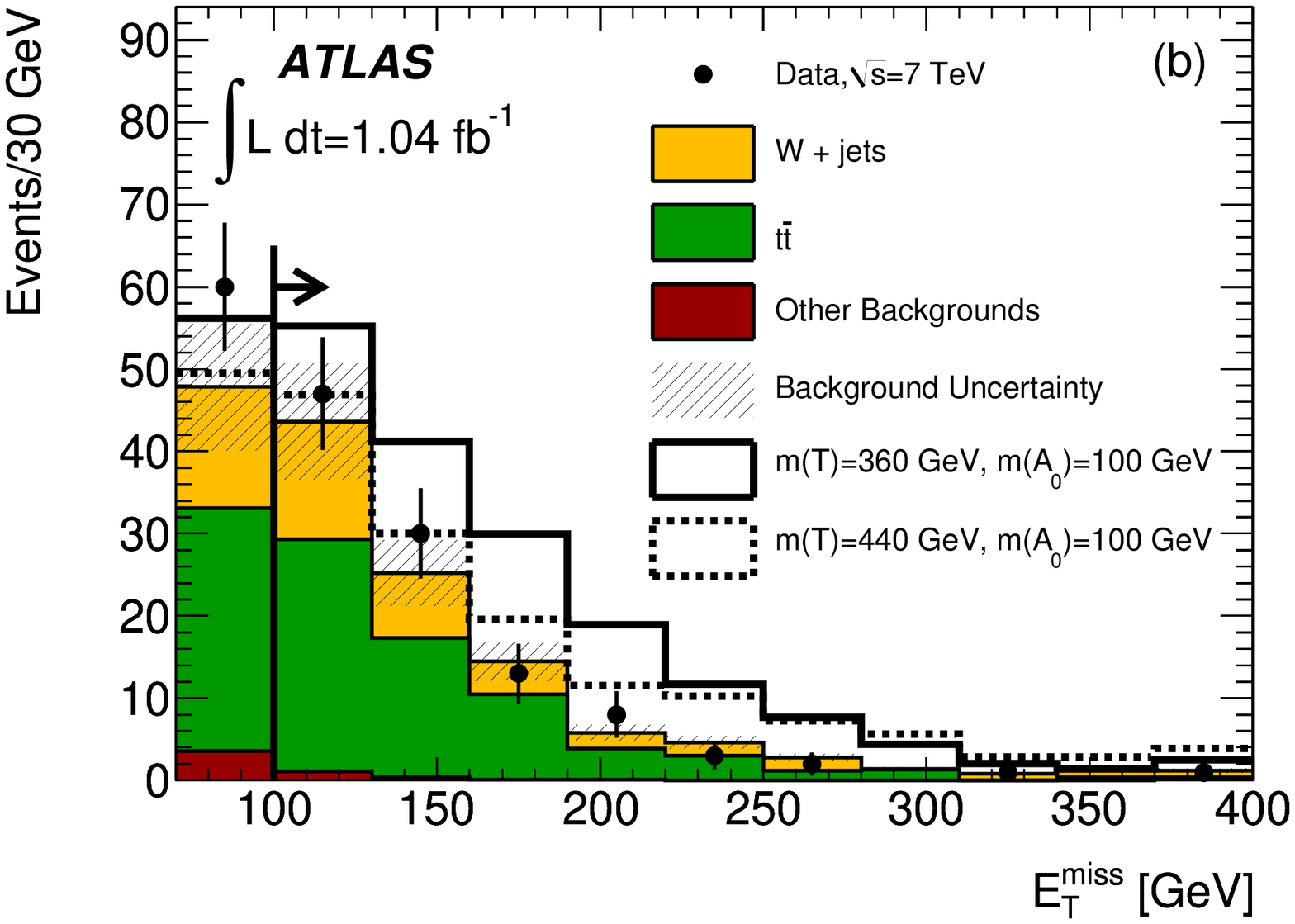}
\includegraphics[width=0.4\textwidth]{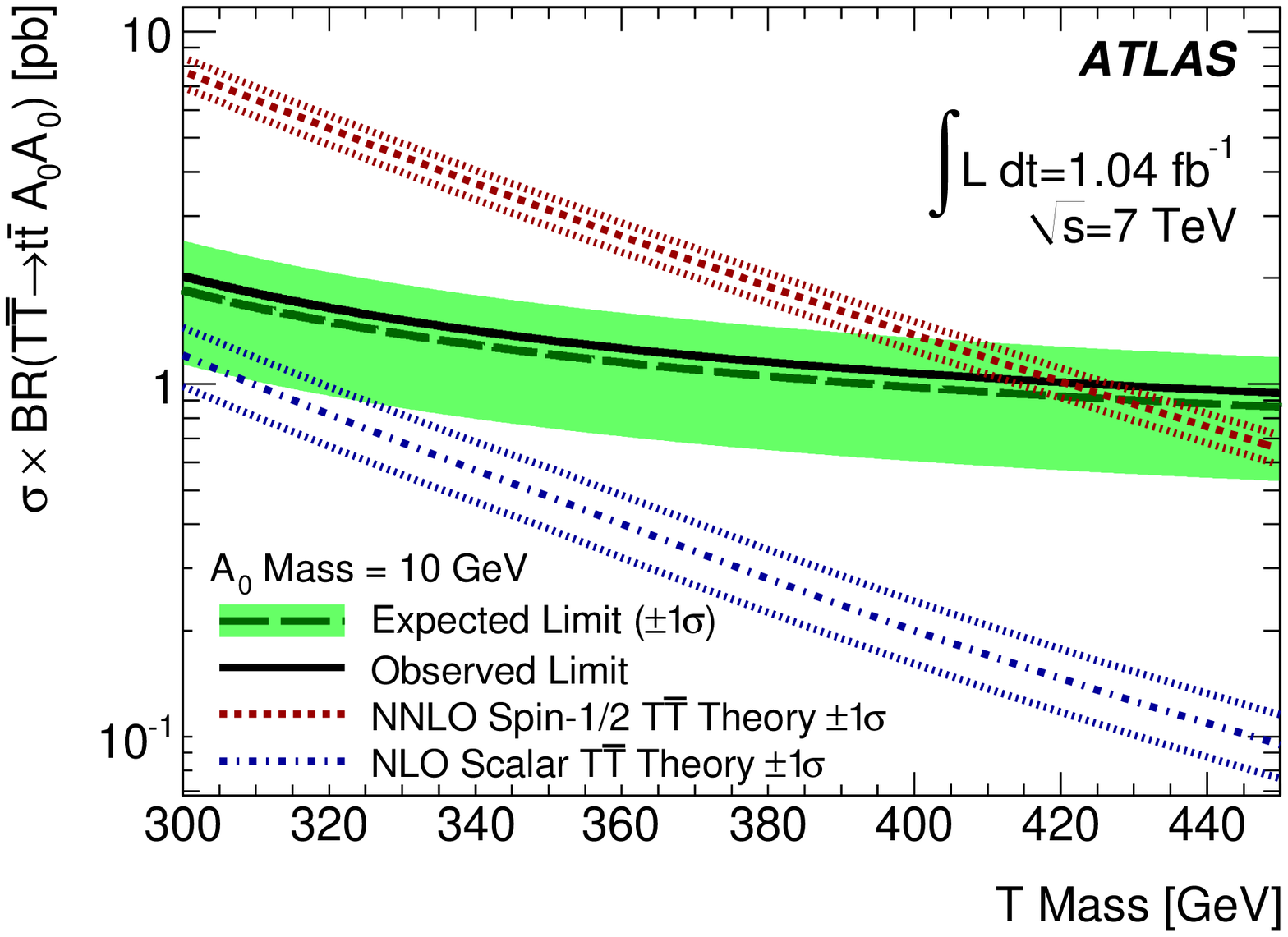}
\end{center}
\caption{Left: $E^{miss}_{T}$ in the search for pair-produced exotic top partners $T\overline{T}$. Right:  $\sigma B$ excluded at the 95\% confidence level versus $T$ mass for $m(A_{0})=$10 GeV. }
\label{fig:TT}
\end{figure}

The top quark holds great  promise as a probe for new phenomena at the TeV scale. It has the strongest 
coupling to the Standard Model Higgs boson, and as a consequence it is the main contributor to the quadratic 
divergence in the Higgs mass. Thus, assuming the ÒnaturalnessÓ hypothesis of effective QFT, light top partners 
(with masses $\sim$1TeV) should correspond to one of the most robust predictions of solutions to the hierarchy problem. 
There are quite a lot of searches for new physics in ATLAS involving top quarks, some of them are presented at this  conference \cite{ilaria}.
As an example, a search for pair-produced exotic top partners $T\overline{T}$, is described \cite{ttbarmet}. Each of these particles decay to top quark and a
neutral weakly-interacting particle ($T\overline{T}\rightarrow t\bar{t}A_{0}A_{0}$) which in some models maybe its own antiparticle.
The final state for such a process is identical to $t\overline{t}$, though with a larger amount of $E^{miss}_{T}$ from undetectable $A_{0}$.
The search is performed in the semileptonic channel, resulting in a final state with an isolated lepton, four or more jets, and high $E^{miss}_{T}$.
The observed yield in the signal region  is compared with the SM expectation, Fig \ref{fig:TT}. In the absence of signal an upper limit on $\sigma B$
is computed. In the model of exotic fourth generation up-type quarks the $T\overline{T}$ production cross-section is predicted to be approximately six 
times higher than for stop squarks with a similar mass, due to the multiple spin states of two $T'$s compared to  scalar stops. 
For this model the cross-section limits are  converted to an exclusion curve in the $T$ vs $A_{0}$ mass parameter space.
No excess is found over  SM predictions, with derived limit $m_{T}>$420 GeV for $m_{A_{0}}>$10 GeV, Fig \ref{fig:TT}.
It should be noted that the estimated acceptance times efficiency for spin-$\frac{1}{2}$ $T\overline{T}$ models is consistent within
systematic uncertainties with that for scalar models, such as pair production of stop squarks. The cross-section limits are 
therefore approximately valid for such models, although the predicted cross-section is typically below the current sensitivity. 

 \section{Summary and Prospects}
 
In this conference report, a subset of ATLAS searches for new phenomena, other than those based on 
Supersymmetry, are brießy described. There is unfortunately no sign of new physics yet at the LHC, 
though most of the limits supersede the ones from the Tevatron experiments. ATLAS however will 
continue to explore a wide variety of signatures, increasing the mass reach and present the results in as 
model-independent manner as possible. In 2012, O(10) fb$^{-1}$ of data is foreseen, with expected increase in 
collision energy to 8 TeV, which will result in signiÞcant gains in mass reach for many BSM phenomena. 
More signiÞcant gains might be expected by pushing towards smaller couplings, compared to the ones in 
common ÕbenchmarkÕ models.


\end{document}